\documentclass{Interspeech2024}

\interspeechcameraready 
\usepackage{graphicx}
\usepackage{caption}
\usepackage{float}
\usepackage{wrapfig}  

\title{Voice Cloning for Dysarthric Speech Synthesis: Addressing Data Scarcity in Speech-Language Pathology}

\name[affiliation={1}]{}{Birger Moëll}
\name[affiliation={2,3}]{Fredrik Sand Aronsson}{}

\address{
  $^1$Division of Speech, Music and Hearing, KTH Royal Institute of Technology, Stockholm, Sweden\\
  $^2$Division of Speech and Language Pathology, Karolinska Institutet, Stockholm, Sweden }
\email{bmoell@kth.se, fredrik.sand@ki.se}


\begin{document}

\maketitle

\begin{abstract} This study investigates the use of voice cloning technology to generate synthetic speech that accurately replicates the unique speech patterns of individuals with dysarthria. Utilizing the TORGO dataset, which comprises speech data from speakers with dysarthria, we aimed to address challenges in data scarcity and privacy that hinder the development of robust models in speech-language pathology. Our primary contributions include: demonstrating the effectiveness of voice cloning in preserving dysarthric speech characteristics; showcasing the utility of speech-to-speech voice cloning techniques; identifying differences between real and synthetic data; and discussing the implications of synthetic data and voice cloning for improving diagnostics, rehabilitation, and functional communication in dysarthria.

We cloned voices of both individuals with dysarthria and controls using a commercial voice cloning platform, ensuring gender-matched synthetic voices to reflect the natural diversity within the dataset. A licensed speech-language pathologist (SLP) evaluated a subset of the recordings for indicators of synthetic speech, dysarthria, and speaker gender. The SLP correctly identified dysarthria in all samples and accurately determined speaker gender in 95\% of cases. Notably, the SLP misclassified 30\% of the synthetic speech samples as real, indicating a high degree of realism in the cloned voices.

Our findings suggest that synthetic speech can effectively capture disordered speech characteristics, and that voice cloning technology has advanced sufficiently to produce high-quality synthetic data that closely resembles real speech, even to trained professionals. This advancement holds significant implications for the health sector where synthetic data can alleviate issues related to data scarcity, privacy concerns, and the need for diverse datasets. By enabling the creation of abundant, high-quality synthetic speech data, voice cloning can enhance AI-driven diagnostics, facilitate the development of generalizable models with acoustic and linguistic markers, personalize therapeutic interventions, and improve assistive communication technologies for individuals with dysarthria.

In conclusion, this study demonstrates the transformative potential of voice cloning and synthetic data in healthcare. We release our synthetic dataset publicly to encourage further research and collaboration, aiming to advance the development of robust, generalizable models that can improve patient outcomes in speech and language pathology. \end{abstract}

\section{Introduction}
Speech generation and voice cloning technologies have undergone significant evolution, transitioning from simple text-to-speech (TTS) systems to sophisticated solutions powered by deep learning \cite{klatt1987review, arik2018neural}. This study explores the utilization of voice cloning technology to produce synthetic speech that accurately replicates the speech characteristics of individuals affected by dysarthria. Our main contributions include: \begin{itemize} \item Demonstrating the effectiveness of voice cloning in preserving the unique speech patterns associated with dysarthria. \item Showing that speech-to-speech voice cloning is an effective techinque in preserving dysarthric speech patterns. \item Highlighting what differences exists between real and synthetic data. \item Discussing how synthetic data and voice cloning can improve diagnostics, rehabilitation, and functional communication for individuals with dysarthria. \end{itemize} The remainder of this paper is structured as follows: Section 2 provides background information on speech differences in dysarthria and an overview of synthetic speech. Section 3 describes our methodology, including the voice cloning model used and evaluation metrics. Section 4 presents our results, followed by a discussion in Section 5, and concludes with outlines for future research.

\section{Dysarthria and synthetic speech}

\subsection{Dysarthria} 
Dysarthria is a motor speech disorder resulting from impaired control of the muscles involved in speech production, which can affect respiration, phonation, articulation, resonance, and prosody. It is typically associated with damage to the central or peripheral nervous systems, including the motor cortex, basal ganglia, cerebellum, brainstem, and cranial nerves. Dysarthria can be a consequence of several neurological conditions with the clinical presentation varying depending on the etiology. Subtypes of dysarthria include flaccid, spastic, ataxic, hypokinetic, hyperkinetic and mixed forms, famously described by Darley, Aronson and Brown in 1975 \cite{darley1975motor}.

\subsection{ Dysarthria in Amyotrophic Lateral Sclerosis}
Amyortrophic Lateral Sclerosis (ALS) affects several different aspects of speech, including imprecise articulation, reduced speech rate, decreased intensity, hypernasality, and reduced intelligibility \cite{Green_2013, tomik2010dysarthria}. Recent advancements in speech analysis and machine learning have shown potential for monitoring the progression of speech impairments in ALS over time. Techniques such as automated speech recognition and acoustic analysis can provide objective measures of changes in speech production longitudinally, which can potentially be used for early diagnosis and the evaluation of treatment efficacy \cite{gutz2019early, an2018automatic, nevler2020automated}. Additionally, voice banking, a process in which individuals record their own speech to create a personalized synthetic voice, can provide a means of communication when speech becomes severely impaired \cite{yamagishi2012speech, Bunnell_2017}.

\subsection{Dysarthria in Cerebral Palsy}

In Cerebral Palsy (CP), dysarthria is caused by early brain damage affecting the motor pathways. The type of dysarthria in CP depends on the motor subtype of CP and commonly includes spastic dysarthria, which is characterized by slow, effortful speech and imprecise articulation due to increased muscle tone \cite{hustad2010classification, scholderle2021dysarthria, scholderle2016dysarthria}. Speech-language pathologists are integral to the multidisciplinary care team, focusing on individualized rehabilitation aimed at improving speech intelligibility, communication efficiency, and quality of life. Emerging technologies, including automated speech analysis, show promise for early identification based on speech samples \cite{allison2018acoustic}.

\subsection{Synthetic Speech}
The advent of synthetic speech data has emerged as a pivotal development in the field of speech technologies, offering a versatile solution to the challenges of data scarcity and privacy concerns. Synthetic data, artificially generated rather than obtained by direct measurement, provides a means to augment datasets, improve model robustness, and facilitate the development of speech technologies without compromising individual privacy \cite{nikolenko2021synthetic, rossenbach2020generating}.
Synthetic data enables the creation of diverse datasets that can capture a wide range of vocal characteristics such as accents, emotions or speech impairments. This diversity can be helpful for training robust models capable of generating natural-sounding speech or transcribing spoken text \cite{kim2020emotional}. Furthermore, synthetic speech data plays a critical role in developing voice cloning technologies, allowing for the creation of personalized synthetic voices with limited real-world samples, thereby preserving the speaker's privacy and identity \cite{arik2018neural}.
The application of synthetic speech in the context of neurological disorders such as ALS is promising. By leveraging voice cloning technologies, it is possible to create personalized synthetic voices that capture the unique speech characteristics associated with these conditions. This approach can not only aid in the development of assistive technologies and communication devices but also contribute to a deeper understanding of the underlying speech impairments \cite{rudzicz2013adjusting, Yamagishi_2012}.
Moreover, synthetic speech can serve as a valuable tool for research and clinical practice. By generating large datasets of synthetic speech samples that mimic the speech patterns of individuals with neurological disorders, researchers can develop and validate new assessment and intervention strategies without relying on real-world data, which may be limited or difficult to obtain \cite{Vachhani_2018, jiao2018simulating}. This can accelerate the development of personalized treatment approaches and improve outcomes for individuals with speech impairments.

\section{Method}

\subsection{Data Collection} For this study, we utilized the TORGO database, a publicly available dataset that provides speech data from speakers with CP and ALS, as well as controls without dysarthria. The TORGO database was developed through a collaboration between the departments of Computer Science and Speech-Language Pathology at the University of Toronto and the Holland-Bloorview Kids Rehab hospital in Toronto \cite{Rudzicz_2012_TORGO}.

The TORGO dataset aims to support the development of advanced automatic speech recognition (ASR) models tailored for dysarthric speech. To achieve this, the dataset provides detailed physiological information, enabling researchers to explore 'hidden' articulatory parameters through statistical pattern recognition. The database includes aligned acoustic data and 3D articulatory features captured using a 3D AG500 electromagnetic articulograph (EMA) system, which records detailed articulatory movements both inside and outside the vocal tract. This setup allows for a comprehensive analysis of speech-related activity in individuals with motor speech disorders \cite{Rudzicz_2012_VocalTract}.

Speakers in the TORGO dataset performed a range of tasks designed to capture various aspects of speech production. These tasks include non-words, short words, restricted sentences, and unrestricted sentences: \begin{itemize} \item \textbf{Non-words}: These tasks are used to gauge the articulatory control of speakers with dysarthria, focusing on phonetic contrasts around plosive consonants and the use of prosody. \item \textbf{Short words}: This category includes repetitions of English digits, simple commands, words from standard dysarthria assessments (e.g., Frenchay Dysarthria Assessment), and phonetically contrasting word pairs. These tasks are valuable for studying speech acoustics without the need for word boundary detection. \item \textbf{Restricted sentences}: These syntactically correct sentences include phoneme-rich phrases and passages designed to utilize lexical, syntactic, and semantic processing in ASR models. \item \textbf{Unrestricted sentences}: To represent naturally spoken speech more accurately, participants were asked to describe images spontaneously, providing a dataset that includes disfluencies and syntactic variations typical of everyday speech. \end{itemize}

Data in the TORGO database is organized by speaker and session, with each speaker assigned a unique code indicating their gender, dysarthric status, and order of recruitment. The dataset includes multiple recording sessions per speaker, containing files related to articulation, phonemic transcriptions, and acoustic data recorded through both array and head-worn microphones. Notably, the articulatory data includes the position, velocity, and orientation of sensor coils placed at key locations on the vocal tract, such as the tongue, lips, and jaw. These data provide a detailed window into the articulatory processes underlying dysarthric speech \cite{Yunusova_2009}.

\subsection{Voice Cloning Model}
We used the commercial voice cloning platform ElevenLabs for voice cloning. 
The systems were chosen due to their state-of-the-art performance and ability to generate high-quality synthetic speech with limited training data.

\subsection{Speech to Speech Voice Cloning}
We used ElevenLabs for Speech to Speech Voice Cloning. Speech to speech voice cloning relies on having a synthetic voice and a speech sample as a prompt for the speech that will be generated. The model generates the content of the speech file in the synthetic voice provided. We used readily available voices in ElevenLabs to transfer the voice to.

\subsection{Voice Cloning} We cloned the entire TORGO dataset to create synthetic voices for each speaker. Each speaker was assigned a unique synthetic voice that matched their gender, ensuring that the synthetic voices reflected the natural diversity within the dataset. The dataset includes three female speakers with dysarthria, five male speakers with dysarthria, three female control speakers, and four male control speakers.

\begin{table}[h!]
\centering
\begin{tabular}{|c|c|c|}
\hline
\textbf{Speaker Group}      & \textbf{Gender} & \textbf{Number of Speakers} \\ \hline
Dysarthria                     & Female          & 3                           \\ \hline
Dysarthria                     & Male            & 5                           \\ \hline
Control                     & Female          & 3                           \\ \hline
Control                     & Male            & 4                           \\ \hline
\end{tabular}
\caption{Distribution of speakers in the synthetic TORGO dataset, matches the original dataset}
\label{tab:speaker_distribution}
\end{table}

For voice cloning, we used the audio from the wav\_headMic recordings when available, as these recordings generally provided higher quality audio. When the wav\_headMic recordings were unavailable, we used the wav\_arrayMic recordings. To ensure the best possible quality for the voice cloning process, we listened to all sessions and selected the recordings with the highest sound quality for each speaker. This helped us create synthetic voices that closely matched the characteristics of the original speakers. By cloning the entire dataset and matching the synthetic voices to the speakers, we tried to preserve the unique speech patterns of individuals with dysarthria as well as the control group.

\subsection{Clinical ratings by SLP}
A subsample of the recording was analysed by the co-author, a licensed speech-language pathologist (SLP). Twenty sessions where chosen with 10 samples for each session with a total of 200 recordings. For each case, an overall assessment was made and the overall rating was used for the results. The data was analyzed based on the following:

\begin{itemize}
    \item \textbf{Synthetic Speech}:Evaluation if the speech sample was real or synthetic
    
    \item \textbf{Dysarthria}: Evaluation if the speech sample contained dysarthria

    \item \textbf{Gender}: Evaluation of the gender of the speaker

    \item \textbf{Quality of Information}: How helpful the information in the audio file was for the classification of gender, dysarthria and synthetic. Rated on a scale 0-3.

    \item \textbf{Audio Quality}: Whether the audio quality was good enough to be used for classification
\end{itemize}

\section{Results}

\begin{table}[ht]
\centering
\caption{Sample Distribution and Results of Speech Analysis by Licensed SLP}
\begin{tabular}{@{}lcc@{}}
\toprule
\textbf{Category} & \textbf{Count} \\ \midrule
\textbf{Real Samples} & 5 \\
\textbf{Synthetic Samples} & 15 \\
\textbf{Aphasia Samples} & 12 \\
\textbf{Control Samples} & 8 \\
\textbf{Female Samples} & 10 \\
\textbf{Male Samples} & 10 \\ \midrule
\textbf{Correct Predictions (Synthetic Speech)} & 14/20 \\
\textbf{False Positives (Synthetic Speech)} & 0 \\
\textbf{False Negatives (Synthetic Speech)} & 6 \\
\textbf{Accuracy (Synthetic Speech)} & 70.00\% \\ \midrule
\textbf{Correct Predictions (Dysarthria)} & 20/20 \\
\textbf{Accuracy (Dysarthria)} & 100.00\% \\ \midrule
\textbf{Correct Predictions (Gender)} & 19/20 \\
\textbf{False Positives (Gender)} & 1 \\
\textbf{Accuracy (Gender)} & 95.00\% \\ \bottomrule
\end{tabular}
\end{table}


\begin{figure}[ht]  
    \hspace{-1cm}  
    \includegraphics[width=0.57\textwidth]{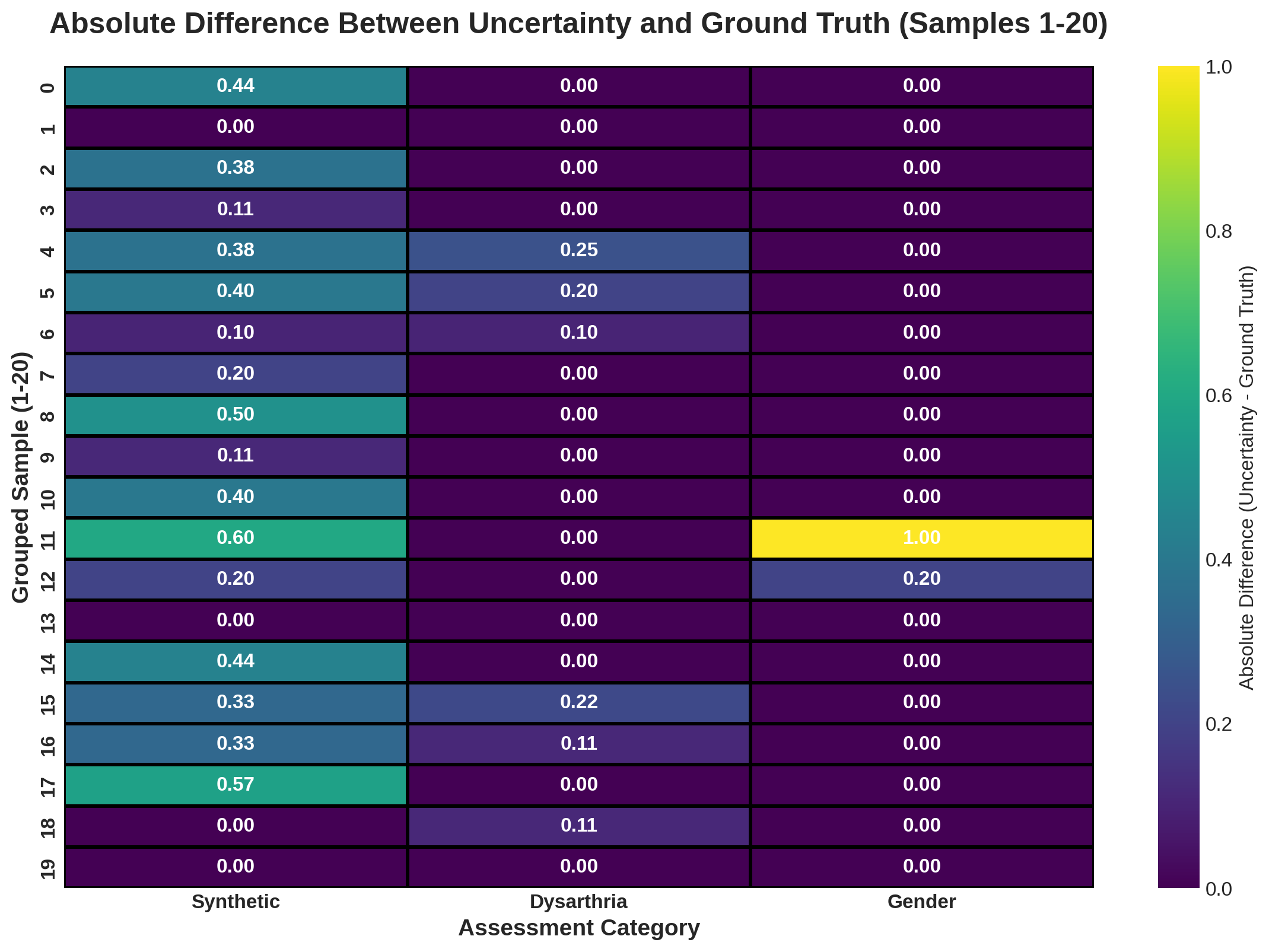}  
    \caption{Difference Between Uncertainty and Ground Truth for Samples 1-20}
    \label{fig:heatmap}
\end{figure}

\subsection{Synthetic Speech Identification}

The SLP’s analysis showed an overall accuracy of 70.00\% in identifying synthetic speech when compared to the ground truth. A detailed breakdown of the errors revealed the following:

Correct Predictions: 14 out of 20 samples were correctly classified as either synthetic or not synthetic.
False Positives: No instances were observed where a non-synthetic sample was incorrectly classified as synthetic.
False Negatives: 6 out of 20 samples (30.00\%) were synthetic but were classified as non-synthetic by the SLP. These errors represent cases where synthetic speech went undetected.
\subsection{Dysarthria Detection}

In terms of dysarthria detection, the SLP’s analysis achieved 100\% accuracy, correctly identifying dysarthria in all cases. This indicates a complete agreement between the SLP’s assessment and the ground truth data for dysarthria.

The fact that a speech-language pathologist misidentified 30\% of the synthetic samples as real is promising and is an indication of the quality of the synthetic speech. The fact that dysarthria detection is 100\% shows that dysarthria is present in the synthetic recordings.

\subsection{Gender Detection}

The accuracy of gender detection was 95.00\%. There was one sample where the gender was misidentified as the opposite gender.

\subsection{Uncertainty}
Figure 1 visualizes the average predictions for all samples relative to the ground truth, providing information on the uncertainty of the Speech Language Pathologist's (SLP) predictions. This uncertainty is calculated based on the variability between the multiple predictions made for the same session.

Synthetic Speech Identification:
In the synthetic speech identification column, we observe varying predictions for many samples. This variability reflects the SLP’s uncertainty in determining whether speech is synthetic, as evidenced by the fluctuating predictions for the same case across multiple attempts. This indicates that there was notable uncertainty regarding whether the speech was generated synthetically or not. For some sessions, the predictions shift between "synthetic" (1) and "non-synthetic" (0), emphasizing the challenges in reliably distinguishing synthetic speech.

Dysarthria Detection:
In the dysarthria column, there is generally more consistency in predictions, with some notable exceptions where predictions diverge. These differences highlight areas where the SLP was uncertain about whether dysarthria was present. This variability in some cases might suggest that the dysarthria in those particular samples was less obvious, leading to differences in judgment across the same sample.

Gender Classification:
The gender classification column shows a high degree of certainty in predictions. There was only one incorrect prediction across all the samples, and for that case, the assessment of gender was consistent across all predictions. This suggests that, in general, gender classification was not a source of significant uncertainty for the SLP.

\section{Discussion}
Our results are promising and in line with a trend happening broadly within generative AI. Humans struggle to identify synthetic samples, including domain experts. Our experiment  demonstrates the feasibility of creating high-quality synthetic data in a complex domain but also underscores the transformative potential of synthetic data in healthcare in general. The last years have seen rapid improvements in generative AI and as models improve so will the ability to create highly realistic synthetic data. 

Collecting real-world data from individuals with speech impairments poses ethical, logistical, and privacy challenges. Synthetic data generation addresses these issues by providing an abundant source of representative speech samples without compromising patient confidentiality. This abundance has the potential to enable the development of more accurate diagnostic tools and personalized therapeutic interventions, ultimately enhancing patient outcomes. This could be especially relevant in rare disorders such as ALS where the number of persons living with the disease might be a limiting factor on the amount of data that can be collected. With the use of synthetic data, the data collection could shift toward a relatively small number of participants where the collected speech samples can form the basis for the creation of a larger synthetic dataset. In the case of diagnostics, synthetic data can enhance the capabilities of artificial intelligence (AI) systems. AI models trained on synthetic datasets can improve the detection of subtle patterns associated with early stages of diseases, leading to earlier interventions. For instance, in neurodegenerative disorders where early diagnosis is critical, synthetic speech data could help in identifying vocal biomarkers that precede more apparent symptoms. The use of synthetic data is of course not limited to synthetic speech, and the implications for the broader health sector are substantial. Synthetic data can alleviate the scarcity of quality data across various medical domains, especially for rare conditions where patient data is limited.

Synthetic data generation also aligns with the stringent privacy regulations like HIPAA and GDPR by minimizing the need to handle sensitive patient information. This reduces the risk of data breaches and can help to build public trust in the use of AI and machine learning in healthcare. Ethical compliance becomes more manageable when synthetic data stands in for real patient data, facilitating broader collaborations and data sharing among researchers and institutions.

The utilization of synthetic data also promotes inclusivity and diversity in healthcare solutions. Real-world datasets often suffer from biases due to underrepresentation of certain demographics. Synthetic data can be engineered to balance these disparities, ensuring that AI models perform equitably across different populations. This is particularly important in speech-language pathology, where dialects, accents, and language variations significantly impact communication patterns.

There are also advantages from an economic perspective. The use of synthetic data can potentially lead to substantial cost savings. Data collection and annotation are resource-intensive processes. By generating synthetic datasets, organizations can reduce expenses related to data acquisition and focus resources on model development and clinical integration. This efficiency accelerates the innovation cycle, bringing advanced healthcare technologies to market more quickly.

In conclusion, the creation of synthetic speech data for individuals with dysarthria is a small but signficant illustration of the broader potential of synthetic data in healthcare. It addresses critical challenges in data availability, privacy, and bias, while presenting new possibilities for personalized medicine and AI-driven diagnostics. As our knowledge of the use of synthetic data advances, it is imperative to continue refining synthetic data technologies, establish ethical frameworks, and develop interdisciplinary collaborations to fully realize the potential of the technology. Embracing synthetic data is not just a technical progression but a strategic imperative that could reshape the future of healthcare delivery and patient care.

The success of the voice cloning highlights the potential benefits of using synthetic data. This pushes speech across the barrier to completely synthetic which in theory allows the creation of unlimited amounts of clinical speech data. Research shows that synthetic data in text, images and speech can improve classification accuracy of models
\cite{chatterjee2022enhancement, venkatesh2021artificially, tang2023does} and that synthetic data can be helpful in the medical domain. \cite{murtaza2023synthetic}. We hope that our data can be helpful in similar ways and we release our dataset freely available online.

\subsection{Limitations}
The small sample size limits the conclusions we can draw from our study. Including only English speakers limits the practical utility of our dataset. Furhtermore, even though the expert rater had substantial difficulties differentiating real from synthetic speech, the samples used sometimes consisted of isolated words which was harder to evaluate in comparison to sentences which provides more information. Additionally, some of the synthetic speech samples, particularly in the sentences, contained atypical linguistic errors similar to paraphasias which disclosed the synthetic nature of the sample. That is, some cases were unintentionally identified as synthetic based on atypical lingusitic features rather than the speech characteristics. These changes aided in the identification and a corrected dataset without linguistic errors would more than likely lower the number of correctly classified cases even further.

\subsection{Future research}
Future research should explore the effectiveness of voice cloning across a broader range of speech and language impairments. A natural next step is to create multi-lingual speech data which could potentially enable the transfer of speech changes across languages, as well as including more dysarthria. We would also like to explore creating classification algorithms solely from synthetic data.

\bibliographystyle{IEEEtran}
\bibliography{mybib}

\end{document}